\documentclass[%
 reprint,
 amsmath,amssymb,
 aps,showpacs
]{revtex4-1}

\usepackage{graphicx}
\usepackage{dcolumn}
\usepackage{bm}

\begin{document}

\title{The dynamics of simple gene network motifs subject to extrinsic fluctuations}

\author{Elijah Roberts$^1$}\thanks{email address: \texttt{erober32@jhu.edu}; Corresponding author}\author{Shay Be'er$^2$}\author{Chris Bohrer$^1$}\author{Rati Sharma$^1$} \author{Michael Assaf$^2$}\thanks{email address: \texttt{assaf@phys.huji.ac.il}; Corresponding author}

\affiliation{$^1$Department of Biophysics, Johns Hopkins University, Baltimore, MD 21218, USA\\
$^2$Racah Institute of Physics, Hebrew University
of Jerusalem, Jerusalem 91904, Israel}

\begin{abstract}
Cellular processes do not follow deterministic rules; even in identical environments genetically identical cells can make random choices leading to different phenotypes. This randomness originates from fluctuations present in the biomolecular interaction networks. Most previous work has been focused on the intrinsic noise (IN) of these networks. Yet, especially for high-copy-number biomolecules, extrinsic or environmental noise (EN) has been experimentally shown to dominate the variation. Here we develop an analytical formalism that allows for calculation of the effect of EN on gene expression motifs. We introduce a new method for modeling bounded EN as an auxiliary species in the master equation. The method is fully generic and is not limited to systems with small EN magnitudes. We focus our study on motifs that can be viewed as the building blocks of genetic switches: a non-regulated gene, a self-inhibiting gene, and a self-promoting gene. The role of the EN properties (magnitude, correlation time, and distribution) on the statistics of interest are systematically investigated, and the effect of fluctuations in different reaction rates is compared. Due to its analytical nature, our formalism can be used to quantify the effect of EN on the dynamics of biochemical networks and can also be used to improve the interpretation of data from single-cell gene expression experiments.
\end{abstract}

\maketitle


\section{Introduction}
Biochemical processes in cells are inherently noisy, because many molecular species such as genes, RNAs, and proteins that make up intracellular reaction networks are present in low copy numbers inside a cell, see \textit{e.g.} Refs.~\cite{Paulsson2004sun,Kaern2005sge}. One of the primary insights to emerge from studies on stochastic gene expression is the distinction between {\it intrinsic} noise (IN) and \textit{extrinsic} noise (EN) \cite{Elowitz2002sge,Swain2002iec,Volfson2006oev,Shahrezaei2008cef,Stewart-Ornstein2012cnr,Hornung2012nrm}. Experimentally, EN is quantified using the correlation in fluctuations between two copies of an identical reporter gene expressed separately in the same cell. IN arises from fluctuations that are independent for each reporter \cite{Hilfinger2011sif}. Within a cell, then, IN is the variance due to the discreteness of biomolecules and the probabilistic nature of chemical reactions, while EN is the variance arising from the fact the genes share a common environment, the cell.

Such noise in cellular reactions can have important consequences, e.g., on cellular decision making. In fact, noise can drive cells between distinct gene expression states corresponding to different decision phenotypes~\cite{Eldar2010frn,Balazsi2011cdm,Golding2011dml}. The ultimate stability of a decision state is then determined by fluctuations of mRNA and proteins, as well as other cellular components, during gene expression~\cite{Hasty2000nsa,Thattai2001ing,Elowitz2002sge,Yu2006pge,Schultz2007mls,Munsky2012uge}. These fluctuations can give rise to spontaneous switching between the states, with a switching time that depends on their strength and the switch's architecture. Genetic switches can regulate diverse decision making processes such as microbial environmental adaptation, developmental pathways, nutrient homeostasis, and bacteriophage lysogeny~\cite{Zeng2010dms,Raj2010vge,Roberts2011nci,Vardi2013bye}.

In recent years there have been numerous theoretical studies on genetic switches driven by IN, noise arising from within a closed system of interest~\cite{Kepler2001str,Aurell2002eaf,Roma2005ope,Hornos2005sge,Mehta2008esn,Morelli2008efr,Bishop2010sbb,Presse2010dfb,Wang2010kpt,Assaf2011dsg,Earnest2013dli,Feng2014gfa,Newby2015bsa,Biancalani2015}.
Yet, these and most other studies of genetic switching, have neglected sources of EN due to interactions with other components in the cell and the environment. EN does not arise from a single well-defined process, but rather results from the complex chain of events that gave rise to a particular cellular state. Variation in a cell's number of ribosomes, transcription factors, and polymerases or fluctuations in the cell division time, as well as environmental fluctuations, can all affect the rates of a genetic process. These fluctuations in the reaction rates may dramatically affect the protein's statistics including its mean, variance and copy-number distribution. Importantly, in living cells a comparison of the relative contribution of EN versus IN to the protein distributions width has shown that EN dominates above copy numbers of ${\cal O}(10-100)$~\cite{Taniguchi2010qec,Sanchez2013gdc,Jones2014pad}.

Theoretically, EN has been shown to induce bistability \cite{Samoilov2005sas,Shahrezaei2008cef,Leisner2009kgs,To2010nci,Caravagna2013}, vary the distribution tails \cite{Shahrezaei2008cef}, and modify switching times \cite{Hu2011ein}. In signaling, EN limits the information transduction capacity of the pathways \cite{Rhee2014ndi,Hansen2015lit}. It has been shown that EN is present at multiple time scales during protein production in bacteria \cite{Hensel2012sed,Singh2014tci} and that negative feedback can filter EN \cite{Hensel2012sed}. Previously, in Ref.~\cite{Assaf2013end} the authors have studied the interplay between IN and EN noise in a genetic switch near bifurcation using a Fokker-Planck approximation and showed that EN can dramatically affect switching. However, their method could not be directly used to study questions regarding population heterogeneity in metastable systems nor with non-Gaussian EN statistics.

Genetic switches and other more complex circuits using multiple positive and negative feedback links form the basis of much of the transcriptional regulatory logic in bacteria. The overall regulatory network of these microorganisms is commonly composed of repeated patterns of relatively small circuits, called motifs. For a review, see \cite{Alon2007nmt}. As a prerequisite to understanding the effect of EN on complex regulatory networks, we would first like to understand the role of EN on simple genetic motifs.

In this work we develop an analytical formalism that allows for the quantification of the effect of EN on intrinsic-noise-driven gene expression circuits. We introduce a new method for modeling bounded EN as an auxiliary species in a master equation that fluctuates according to non-Gaussian statistics. We then analyze three genetic motifs: a non-regulated gene, a self-inhibiting gene, and a self-promoting gene. These three motifs represent the simplest possible circuits, yet commonly occur in bacterial transcription networks. We study the properties of each motif as related to EN. All our analytical findings are tested and compared against numerical Monte Carlo simulations.

\section{Theory}
\subsection{Gene expression under intrinsic noise}
Our starting point is a simple gene-expression model without extrinsic noise (EN). Let us denote by $n$ the protein's copy number and by $N\gg 1$ the typical protein abundance in the steady state. Our gene-expression model will consist of two reactions: production of proteins at a rate of $F(n)$ and degradation with rate $\nu n$. Here we assume that the mRNA lifetime is short compared to the cell cycle and momentarily ignore the mRNA fluctuations, which will be accounted for in the following.

In the deterministic picture, the rate equation for the protein concentration $x=n/N$ reads
\begin{equation}\label{RE}
\dot{x}=f(x)-x,
\end{equation}
where $f(x)=F(n)/N$ and we have rescaled time by the protein degradation rate $\nu$. To account for intrinsic fluctuations due to the probabilistic reactions and the discreteness of the proteins, we write down the chemical master equation for $P_n(t)$ -- the probability to find $n$ proteins at time $t$
\begin{equation}\label{ME}
\dot{P}_n=F(n-1)P_{n-1}+(n+1)P_{n+1}-[F(n)+n]P_n.
\end{equation}
We look for the stationary PDF such that $\dot{P}_n=0$. This yields a set of recursive equations, whose solution can be found analytically. The solution reads~\cite{Gardiner2004hsm}
\begin{equation}\label{PDFIN}
P_n=P_0\prod_{m=0}^{n-1}\frac{F(m)}{m+1}=P_0\exp\left[\sum_{m=0}^{n-1}\ln \frac{F(m)}{m+1}\right],
\end{equation}
where $P_0$ is a normalization factor such that $\sum_{n=0}^{\infty}P_n=1$.

The stationary solution of Eq.~(\ref{ME}) can also be found by using a dissipative WKB approximation~\cite{Bender1999,Dykman1994lfo}. To this end, we assume $n\gg 1$, treat $n$ as a continuous variable, and search for $P_n$ as $P_n\equiv P(x)\sim e^{-NS(x)}$. Here, $N \gg 1$ is assumed to be a large parameter, and $S(x)$ is called the action. Plugging this ansatz into the stationary master equation [Eq.~(\ref{ME}) with $\dot{P}_n=0$], we arrive in the leading ${\cal O}(N)$ order at a stationary Hamilton-Jacobi equation $H[x,S'(x)]=0$ with Hamiltonian $H(x,p_x)=f(x)(e^{p_x}-1)+x(e^{-p_x}-1)$, where we have denoted the associated momentum by $p_x=S'(x)$. While the trivial zero-energy trajectory $p_x(x)=0$ of this Hamiltonian corresponds to the deterministic dynamics, in the leading order the statistics of interest are encoded in the nontrivial zero-energy trajectory of this Hamiltonian~\cite{Dykman1994lfo}, which reads in this case
\begin{equation}\label{momentum}
p_x(x)=\ln[x/f(x)].
\end{equation}
This allows us to calculate the action by integration $S(x)=\int^x p_x(x')dx'$. Thus, the PDF and its variance due to IN, $var_{_{IN}}=NS''(x_*)^{-1}$~\cite{Assaf2010ems}, are found to be:
\begin{equation}\label{PDFWKB}
P(x)\simeq \sqrt{\frac{S''(x_*)}{2\pi N}} e^{-N[S(x)-S(x_*)]},\;\;var_{_{IN}}=\frac{Nx_*}{1-f'(x_*)},
\end{equation}
where $x_*$ is the steady-state solution of Eq.~(\ref{RE}), and the normalization was done over the Gaussian part of the PDF around $x_*$. By transforming the sum into an integral,  Eq.~(\ref{PDFIN}) coincides in the leading order in $N\gg 1$ with the PDF in Eq.~(\ref{PDFWKB}).

\subsection{Gene expression under intrinsic and extrinsic noise}
Next, we add to our model EN, which is commonly defined as intercellular variability due to fluctuations during gene expression that equally affect all genes within a cell. We thus introduce EN in the form of one or more fluctuating parameters. For concreteness we assume, \textit{e.g.}, that cell-to-cell variability in transcription and translation rates causes the protein degradation rate $\nu$ to fluctuate so that $\nu\to \nu(t)=\xi(t)$. (In Appendix C we consider other fluctuating parameters as well.) As a result, the degradation rate becomes $n\xi(t)$ where $\xi(t)$ is a stochastic variable satisfying $\langle \xi(t)\rangle=1$. Many measured protein distributions appear to be well-fit by a negative binomial (or gamma) distribution. Without experimental knowledge of how rates fluctuate \textit{in vivo}, we simply take $\xi(t)$ to have a negative binomial statistics, as if being controlled by a single protein. In addition, $\xi(t)$ has variance $\sigma_{ex}^2$ and correlation time $\tau_c$, satisfying  $\langle \xi(t)\xi(t')\rangle=\sigma_{ex}^2 e^{-|t-t'|/\tau_c}$. Other statistics are also possible~\cite{Shahrezaei2008cef,Caravagna2013}; in Appendix D we consider Ornstein-Uhlenbeck noise. Note that our choice of negative binomial statistics for the EN ensures that the rates are always positive.

To model EN, we need a circuit that generates an auxiliary species whose copy number fluctuates with negative binomial statistics and correlation time $\tau_c$. To create one, we use an auxiliary mRNA-protein circuit where mRNAs are transcribed at a rate $\alpha/\tau_c$ and degrade with rate $\omega/\tau_c$, while proteins are translated at a rate $\omega \beta/\tau_c$ and degrade at a rate $1/\tau_c$, which ensures that the correlation time of the auxiliary proteins is $\tau_c$. As a result, the master equation describing the probability to find $m$ auxiliary mRNAs and $k$ auxiliary proteins satisfies:
\begin{eqnarray}\label{auxmasterMain}
&&\dot{P}_{m,k}=\frac{\alpha}{\tau_c}(P_{m-1,k}\!-\!P_{m,k})+\frac{\omega}{\tau_c}[(m+1)P_{m+1,k}\!-\!mP_{m,k}]\nonumber\\
&&+\frac{\omega \beta m}{\tau_c}(P_{m,k-1}\!-\!P_{m,k})+\frac{1}{\tau_c}[(k+1)P_{m,k+1}\!-\!kP_{m,k}]\!.
\end{eqnarray}

As shown in Appendix A, in the limit of short-lived mRNA such that $\omega\gg 1$, the stationary PDF of the auxiliary protein is~\cite{Paulsson2000rsf}:
\begin{equation}
P_k=\frac{\Gamma(\alpha+k)}{\Gamma(k+1)\Gamma(\alpha)}\left(\frac{\beta}{\beta+1}\right)^k\,\left(\frac{1}{\beta+1}\right)^{\alpha},
\end{equation}
where $P_k$ is the probability to find $k$ auxiliary proteins. Here $k=K\xi$, where $K\equiv\alpha\beta$ is the PDF mean, while the variance is $K(1+\beta)$. Therefore, choosing $\alpha=1/(\sigma_{ex}^2-1/K)$ and $\beta=K\sigma_{ex}^2-1$ we find that $\langle\xi\rangle=\alpha\beta/K=1$ and the variance of $\xi$ becomes $K(1+\beta)/K^2=\sigma_{ex}^2$ as required by our EN stochastic variable. Note, that in the limit of large $K$ such that $\beta=K\sigma_{ex}^2-1\simeq K\sigma_{ex}^2$ and $\alpha\simeq 1/\sigma_{ex}^2$, the negative binomial distribution can be well approximated by a gamma distribution $P_k\simeq \beta^{-\alpha}/\Gamma(\alpha)\,k^{\alpha-1}e^{-k/\beta}$~\cite{Cai2006spe}.

To study the interplay between IN and EN, we combine the EN dynamics [Eq.~(\ref{auxmasterMain})] with the underlying intrinsic noise dynamics [Eq.~(\ref{ME})]. This leads to a 3D master equation describing the evolution of the probability $P_{n,m,k}$ to find $n$ proteins, $m$ auxiliary mRNAs and $k$ auxiliary proteins, where the death rate of the protein of interest depends on the auxiliary protein. To this end, by using the WKB theory and by adiabatically eliminating the short-lived auxiliary mRNA degree of freedom (see Appendix A), we arrive at a stationary Hamilton-Jacobi equation $H=0$ with a reduced Hamiltonian for the protein of interest and auxiliary protein:
\begin{eqnarray}\label{2D-ham}
&&H(x,p_x,\tilde{\xi},p_{\tilde{\xi}})=f(x)(e^{p_x}-1)+\frac{x\tilde{\xi}}{\rho}(e^{-p_x}-1)\nonumber\\
&&+\frac{\rho}{\beta\tau_c}\left[\frac{1}{1+\beta(1-e^{p_{\tilde{\xi}}})}-1\right]
+\frac{\tilde{\xi}}{\tau_c}(e^{-p_{\tilde{\xi}}}-1).
\end{eqnarray}
Here we have defined a rescaled EN variable $\tilde{\xi}=\rho\xi$, where $\rho=K/N$ is the abundances ratio of the auxiliary protein and protein of interest. Also, $p_x$ and $p_{\tilde{\xi}}$ are the momenta associated with the protein of interest and the auxiliary protein with corresponding concentrations $x=n/N$ and $\tilde{\xi}=k/N$, while $\alpha$ and $\beta$ are defined above. Hamiltonian~(\ref{2D-ham}) encodes the stochastic dynamics of a protein when its degradation rate fluctuates due to negative binomial EN generated by another auxiliary protein. Note, that while the (arbitrary) copy number $K$ of the auxiliary protein enters Hamiltonian~(\ref{2D-ham}), it \textit{does not} enter the results below for the statistics of the protein of interest.

Hamiltonian~(\ref{2D-ham}) can be theoretically analyzed by writing down the corresponding Hamilton equations, see Eqs.~(\ref{HEgamma-aux}) in Appendix B. These can be solved numerically for arbitrary $\tau_c$, see Methods section. Analytical progress can be made in two important limits: short-correlated ``white'' EN, and long-correlated ``adiabatic'' EN.

In the white-noise limit, $\tau_c\ll 1$, one arrives at a reduced white-noise Hamiltonian, which effectively takes into account the short-correlated EN. Solving the corresponding Hamilton-Jacobi equation we find (see Appendix B)
\begin{equation}\label{pxgamma-main}
p_x=\ln\left\{\frac{x}{2f(x)}\left[1\!-\!V\tau_c x\!+\!\sqrt{(V\tau_c x\!-\!1)^2+4V f(x)\tau_c}\right]\right\}.
\end{equation}
Here, $V\equiv N\sigma_{ex}^2$ is the ratio between the relative EN and IN variances (where the IN variance is taken in the non-regulated case). From Eq.~(\ref{pxgamma-main}) we can calculate the action $S(x)=\int^x p_x(x')dx'$, while the PDF is given by Eq.~(\ref{PDFWKB}). The action function $S(x)$ cannot be explicitly calculated without specifying $f(x)$. Yet, a general result for the PDF variance can be derived. Differentiating $p_x(x)$ [Eq.~(\ref{pxgamma-main})] once and plugging $x=x_*$ such that $f(x_*)=x_*$, we find the observed PDF variance
\begin{equation}\label{wnvargamma-main}
\sigma_{obs}^2=NS''(x_*)^{-1}=\frac{Nx_*(1+x_*V\tau_c)}{1-f'(x_*)}.
\end{equation}
Comparing with Eq.~(\ref{PDFWKB}), this indicates that short-correlated EN increases the variance by a factor of $1+x_*V\tau_c$.

In the adiabatic limit, $\tau_c\gg 1$, we can assume that the EN is almost stationary~\cite{Assaf2013end}. As a result, the protein PDF can be written as
\begin{equation}\label{adiabaticPDF}
P_n=\int_{-\infty}^{\infty}P(\xi)P(n|\xi)d\xi,
\end{equation}
where $P(n|\xi)$ is the conditional probability to find $n$ proteins given noise magnitude $\xi$, and $P(\xi)$ is the probability to find EN magnitude $\xi$. In Eq.~(\ref{adiabaticPDF}) we effectively optimize the ``cost'' of reaching a state with $n$ proteins given EN magnitude $\xi$ against the probability of choosing such $\xi$~\cite{Assaf2013end}.

For simplicity, here we take gamma-distributed EN, $P(\xi)=\tilde{\beta}^{-\alpha}/\Gamma(\alpha)\,\xi^{\alpha-1}e^{-\xi/\tilde{\beta}}$, where $\alpha\simeq 1/\sigma_{ex}^2$ and $\tilde{\beta}=\beta/K\simeq \sigma_{ex}^2$. This distribution has a mean of $1$ and variance $\sigma_{ex}^2$ as required, and is a good approximation to the negative binomial distribution for large $K$~\cite{Cai2006spe}. Performing the integration in (\ref{adiabaticPDF}) via the saddle-point approximation, see Appendix B, the PDF in the adiabatic limit reads
\begin{equation}\label{adiapdfgam-main}
P(x)\simeq  \frac{C}{\sqrt{\partial_{\xi\xi}\Phi[x,\xi=\xi_*(x)]}}\frac{e^{-N\Phi[x,\xi=\xi_*(x)]}}{\xi_*(x)},
\end{equation}
where
\begin{eqnarray}\label{fungam-main}
\Phi(x,\xi)=\int_{g(\xi)}^x\ln\frac{y\xi}{f(y)}dy+\frac{\xi-\ln\xi-1}{V},
\end{eqnarray}
is the cost function. Here, $\xi_*(x)$ is the solution of the saddle point equation $\partial_{\xi}\Phi(x,\xi)=0$, $x=g(\xi)$ is the $\xi$-dependent stable fixed point found by solving the equation $f(x)=x\xi$, and $C$ is a normalization factor such that $N\int_0^{\infty}P(x)dx=1$.

Similarly as in the white-noise case, here we can also calculate the variance of the PDF explicitly for any production rate $f(x)$. After some algebra, see Appendix B, we find the observed variance of the PDF
\begin{equation}\label{adiavargam-main}
\sigma_{obs}^2=\frac{Nx_*}{1-f'(x_*)}\left[1+\frac{Vx_*}{1-f'(x_*)}\right].
\end{equation}
This indicates that adiabatic EN increases the variance compared to the IN-only case~(\ref{PDFWKB}) by a factor of $1+Vx_*/[1-f'(x_*)]$.

Eqs.~(\ref{wnvargamma-main}) and (\ref{adiavargam-main}) for the variance are among our main results here. When EN is put in the production rate instead of the degradation rate the results for the variance in both the white- and adiabatic-EN cases remain the same, see Appendix C.

\section{Results}
\begin{figure}[h]
\begin{center}
\includegraphics{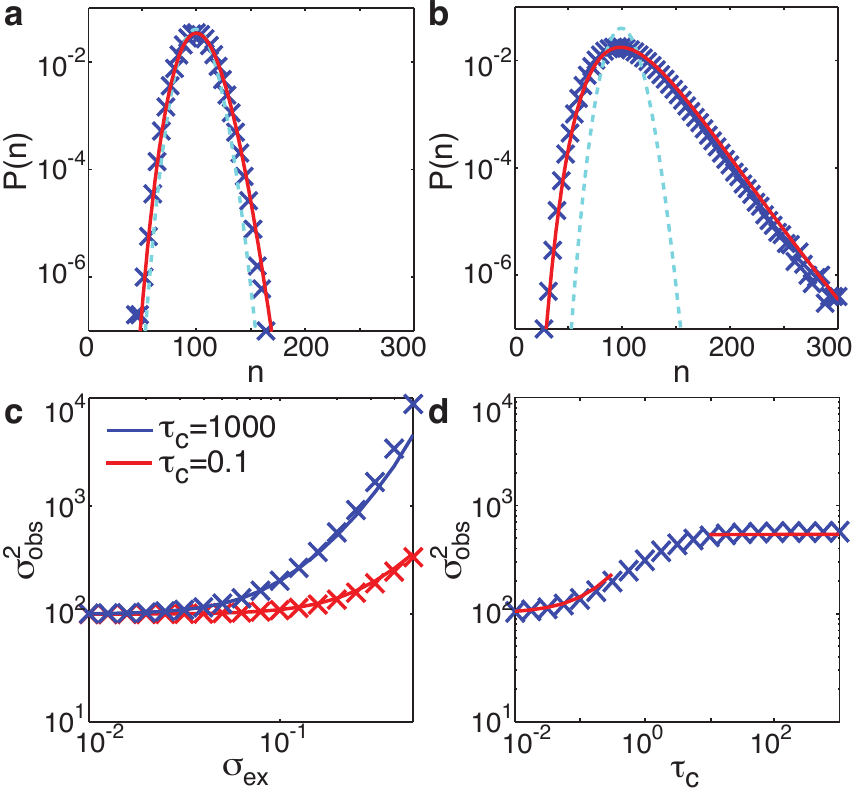}
\end{center}
\caption{(Color online) Comparison of theory (lines) and numerics (symbols) for the non-regulated gene model with $N=100$ and $\sigma_{ex}=0.2$. (a+b) Probability distributions for white (a; $\tau_c=0.1$) and adiabatic (b; $\tau_c=1000$) EN. Dotted lines show the Poisson distribution for the model with only intrinsic noise. (c) Observed variance vs EN strength for white (red) and adiabatic (blue) noise. (d) Observed variance vs EN correlation time for $\sigma_{ex}=0.2$. The left curve shows the white noise theory and the right shows the adiabatic theory.}
\label{fig:UnregulatedGene}
\end{figure}
\subsection{Unregulated gene expression}
We begin with a model for protein transcription given a constant birth rate, namely an unregulated gene. Here the rate equation is given by Eq.~(\ref{RE}) with $f(x)=1$, while the protein PDF is $P_n=e^{-N}N^n/n!$.

Now, we add EN to the protein's degradation as described above. In the white-noise limit, $\tau_c\ll 1$, integrating over the momentum~(\ref{pxgamma-main}) with $f(x)=1$, the action function becomes $S(x)=1/(V\tau_c)\left\{\Omega(x)+\ln[V\tau_c x-1 + \Omega(x)]\right.$ $\left.+V\tau_c x\left[\ln\left((x/2)(1-V\tau_c x + \Omega(x))\right)-1\right]\right\}$,
with $\Omega(x)=\sqrt{(V\tau_c x-1)^2+4V\tau_c}$. Using $S(x)$, we find the PDF, given by Eq.~(\ref{PDFWKB}), see Figure \ref{fig:UnregulatedGene}(a).
Interestingly, in the presence of EN the far right tail of the PDF behaves as a power law. Indeed, taking the $x\gg 1$ limit of the PDF, we find a power-law dependence in the leading order $P(x)\sim (2V\tau_c x)^{-N/(V\tau_c)}$, in contrast to an exponential tail in the IN-only case. Nevertheless, because the power-law behavior appears only at $x\gg x_*=1$, the corresponding probabilities are vanishingly small and thus, observing this behavior experimentally or even numerically is impractical.

The variance of the PDF  due to white EN  [Eq.~(\ref{wnvargamma-main})] becomes $\sigma_{obs}^2=N(1+V\tau_c)$ (Figure \ref{fig:UnregulatedGene}(c,d)). That is, white EN increases the width of the PDF by a factor of $\sqrt{1+V\tau_c}$.

In the adiabatic limit, $\tau_c\gg 1$, the PDF is given by Eq.~(\ref{adiapdfgam-main}) (Figure \ref{fig:UnregulatedGene}(b)). Here, the cost function [Eq.~(\ref{fungam-main})] becomes $\Phi(x,\xi)=(\xi-\ln\xi-1)/V+\int_{1/\xi}^x\ln(y\xi)dy$, where $x=g(\xi)=1/\xi$ is the solution to the equation $f(x)=\xi x$ with $f(x)=1$. In addition, the saddle point is found at $\xi_*=(1-Vx)/2\left(1+\sqrt{1+4V/(1-Vx)}\right)$, while $\partial_{\xi\xi}\Phi(x,\xi)=(2V +\xi-Vx\xi)/(V\xi^3)$.

Plugging $f(x)=1$ and $x_*=1$ into Eq.~(\ref{adiavargam-main}), the variance due to adiabatic EN is given by $\sigma_{obs}^2=N(1+V)$ (Figure \ref{fig:UnregulatedGene}(c,d)). That is, adiabatic EN increases the width of the PDF by a factor of $\sqrt{1+V}$ which can be significant when $V\gtrsim {\cal O}(1)$.

To test our theory we performed Monte Carlo simulations of the full master equation describing all three species: the protein of interest $n$, the auxiliary mRNA $m$, and the auxiliary protein $k$, see Methods. Figure \ref{fig:UnregulatedGene} shows example comparisons for $N=100$ and $\sigma_{ex}$ ranging up to 0.5, which are typical values obtained from single-cell \textit{Escherichia coli} protein distributions \cite{Taniguchi2010qec}. Good agreement is obtained between our theory and stochastic simulations for both the white and adiabatic cases, even for quite strong EN. We have also verified that the results hold for EN in the birth rate, see Appendix C and Figure S3 in~\cite{SuppInfo}. Interestingly, when EN arises in the degradation term, the PDF mean shifts to the right due to the nonlinear dependence of the fixed point on the death rate. While it is negligible for weak and moderate EN, this shift in the mean becomes significant for very strong EN, when $\sigma_{ex}={\cal O}(1)$, see the end of Appendix B for details. In turn, this shift affects the IN of the system, see Appendix E. In such a case IN and EN cannot be independently separated, as is commonly assumed (see Figure S3 and Figure S4 in~\cite{SuppInfo}).

\subsection{mRNA-protein model with no feedback}
Now we consider the more realistic case of an unregulated gene but with mRNA present in the model. Here mRNAs are transcribed at a rate $a$, decay with a rate $\gamma$, and translation of proteins occurs with rate $\gamma b$ while degradation of proteins occurs with rate $1$. As in the auxiliary circuit, we take the ratio between the mRNA and protein degradation rates to be large $\gamma\gg 1$. The mean protein number here is $N\equiv ab$. The rate equations describing the average mRNA and protein concentrations, $r=l/N$ and $x=n/N$ (with $l$ and $n$ being the respective copy-numbers of mRNA and proteins), are $\dot{r}=a/N-\gamma r$, and $\dot{x}=b\gamma r-x$.

In the limit of short-lived mRNA, $\gamma\gg 1$, the stochastic dynamics has been analyzed by various authors~\cite{Assaf2011dsg}.
Using the WKB approximation one can find the protein PDF, see Appendix A, which coincides with the negative binomial distribution in the limit of $n\gg 1$. In particular, the PDF variance becomes $N(1+b)$, indicating that mRNA noise increases the variance by a factor of $1+b$ compared to the protein-only case~\cite{Thattai2001ing}.

We now proceed to calculate the observed variance of the proteins of interest under negative binomial adiabatic EN in the protein's degradation rate. We do so along the same lines done for the protein-only case. Here, accounting for mRNA noise, the momentum given noise magnitude $\xi$ becomes $p_x(x,\xi)=\ln[(1+b)x\xi/(1+b x\xi)]$, which reduces to the protein-only case for $b\to 0$. Integrating over the momentum, we find the action to be $S(x,\xi)=x\ln\left[(1+b)x\xi/(1+bx\xi)\right]-1/(b\xi)\ln(1+bx\xi)$.
Now, similarly as done in Eq.~(\ref{fungam-main}), we can define the cost function $\Phi(x,\xi)=S(x,\xi)-S[g(\xi),\xi]+(\xi-\ln\xi-1)/V$. Therefore, the variance of the PDF can be found using Eq.~(\ref{secdergam}), by repeating the calculations along the same lines as done in Appendix B for the protein-only case. As a result, we find the observed variance of the proteins of interest, while accounting for mRNA noise, to be
\begin{equation}\label{eq:varianceburst}
\sigma_{obs}^2=N(1+b+V).
\end{equation}

\begin{figure}[h]
\begin{center}
\includegraphics{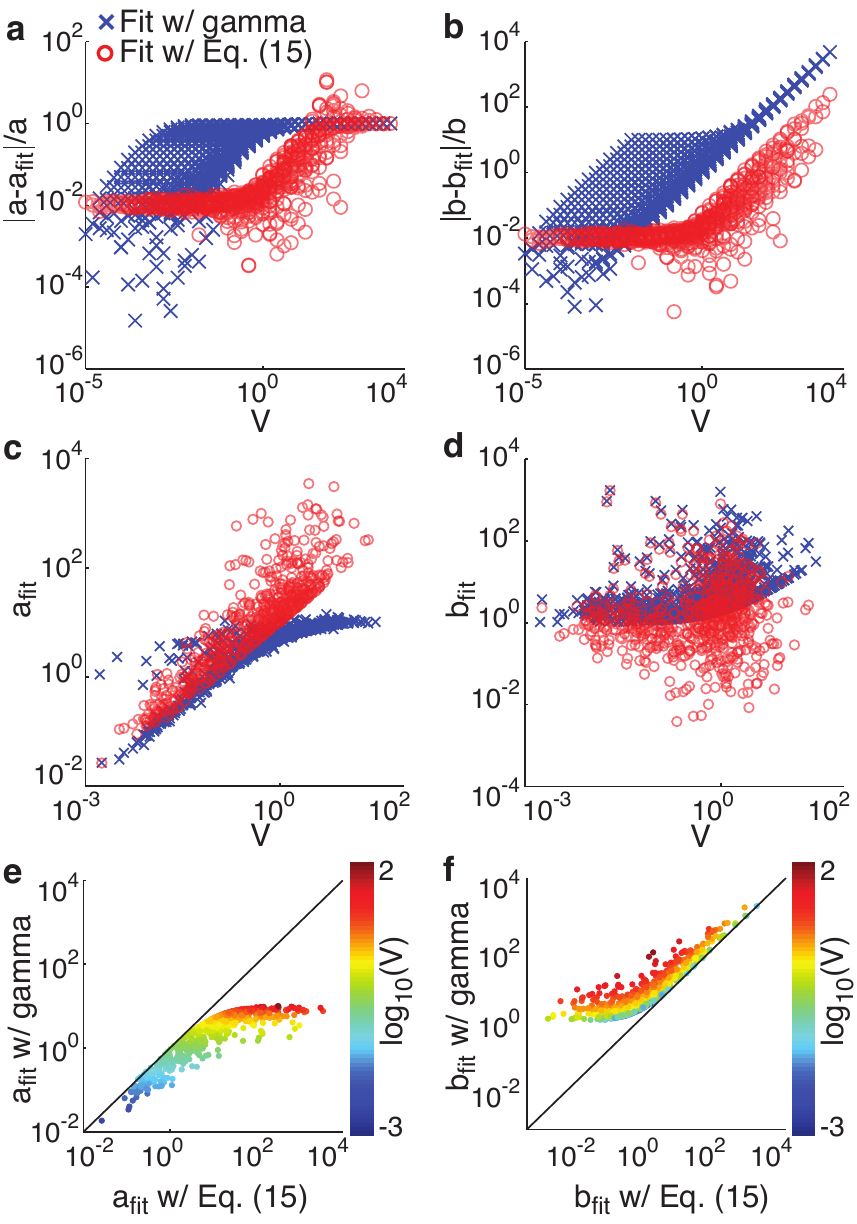}
\end{center}
\caption{(Color online) (a+b) Relative error in $a_{fit}$ and $b_{fit}$, respectively, from simulations spanning a wide range of parameters. Blue crosses show the values from fitting to a gamma distribution. Red circles show fits from Eq.~(\ref{eq:varianceburst}).(c+d) The $a_{fit}$ and $b_{fit}$ values vs $V$ from fitting the genome-scale protein abundance data from \cite{Taniguchi2010qec} using (blue crosses) the gamma distribution or (red circles) Eq.~(\ref{eq:varianceburst}) with a global EN of $\sigma_{ex}=0.31$. (e+f) Relationship between $a_{fit}$ and $b_{fit}$ values obtained from a gamma distribution and Eq.~(\ref{eq:varianceburst}). Points are colored by $log_{10}(V)$ where $V=\sigma^2_{ex}/\sigma^2_{int}$. The solid $y=x$ lines are a guide to the eye.}
\label{fig:BurstModelParameterFits}
\end{figure}

The gamma distribution is widely used to analyze single-cell protein abundance data \cite{Cai2006spe}. The protein's PDF is fit to a gamma distribution and the $a$ and $b$ values resulting from the fit are interpreted as the gene's burst frequency and burst size, respectively. We wanted to study how EN would affect such interpretations. To this end, we performed a large number of stochastic simulations across a wide range of values for $a$, $b$, and $\sigma^2_{ex}$, again for biological ranges seen in single-cell experiments, and calculated stationary PDFs using $\mathrm{10^7}$ data points for each parameter set. We fit the PDFs to a gamma distribution to obtain estimates of the gene expression parameters $a_{fit}$ and $b_{fit}$. To calculate the accuracy with which $a_{fit}$ and $b_{fit}$ recovered the actual parameters, we calculated the relative error as $Err(a) = |a-a_{fit}|/a$ and $Err(b) = |b-b_{fit}|/b$ using the known $a$ and $b$ values from the simulations. As can be seen in Figure \ref{fig:BurstModelParameterFits}(a+b), using a gamma distribution resulted in poor estimates even in the case of relatively weak EN. Given the sensitivity of the error to EN, gamma distribution estimates of gene expression parameters should be used with caution.

We then instead used Eq.~(\ref{eq:varianceburst}) along with the gene-by-gene EN values of $\sigma^2_{ex}$ to estimate $a_{fit}$ and $b_{fit}$ values from the simulated dataset. With the mean of the distribution $N = a\,b$ and the observed variance $\sigma_{obs} =  N(1+b+V) = a \,b (1 + b + a \, b \, \sigma_{ex}^2)$, one can solve directly for $a= N^2/(\sigma_{obs}^2-N-N^2 \sigma_{ex}^2)$ and $b = \sigma_{obs}^2/N-1-N \sigma_{ex}^2$. Using the calculated mean and variance of the PDF, along with the known $\sigma^2_{ex}$ from the simulations, we recovered estimates for $a_{fit}$ and $b_{fit}$. Figure \ref{fig:BurstModelParameterFits}(a+b) shows that if one knows the strength of the EN for a gene, Eq.~(\ref{eq:varianceburst}) can reliably recover the true $a$ and $b$ values until the total variance becomes dominated by EN for $V>10$.

Next we attempted to recover the $a_{fit}$ and $b_{fit}$ values from a genome-scale protein abundance data set from \textit{E. coli} \cite{Taniguchi2010qec}. Here, we made a simplifying assumption that a constant global EN of $\sigma_{ex}=0.31$ influenced all genes equally (see Figure S5 in~\cite{SuppInfo}). We estimated the $a_{fit}$ and $b_{fit}$ values for each gene using the gamma distribution and also using Eq.~(\ref{eq:varianceburst}) with this global EN. Figure \ref{fig:BurstModelParameterFits}(c-f) shows a comparison of the two methods. A few trends are apparent from the results. When accounting for EN, the global saturation in the burst frequency $a$ disappears and instead we see a continuous linear increase in the burst frequency. Likewise, an observed global increase in $b$ values at higher $V$ disappears and a more uniform distribution of $b$ values is seen with respect to $V$. Since $V$ is correlated with overall expression levels (IN goes down as $V$ goes up) this implies that burst frequency is a significant driver of protein expression levels in \textit{E. coli}. Figure S6, see~\cite{SuppInfo}, shows the fits versus mean expression. Gene-by-gene estimates of EN, rather than a single global EN, would lead to even better estimates of gene expression parameters.

\subsection{Self-inhibiting gene}
Next we consider the case of a self-inhibiting gene. A self inhibiting gene is a simple yet common motif that is capable of filtering some types of IN \cite{Dublanche2006ntn}, although it can lose effectiveness when multiple time scales are involved \cite{Radulescu2012rnr}. We were therefore interested to study the ability of a self-inhibiting gene to filter EN.

The rate equation is given by Eq.~(\ref{RE}) where we took the production rate to be $f(x)=(1+\beta)/(1+\beta x)$,
and $\beta$ is the inhibition strength. Here we chose a simple form of nonlinear inhibitory Hill-like function with Hill coefficient $h=1$ (below we will consider higher values of $h$ as well), whose fixed point $x_*=1$ coincides with the non-regulated gene.

To find the PDF in the IN-only case, we integrate over Eq.~(\ref{momentum}) using $f(x)=(1+\beta)/(1+\beta x)$. This yields
$S(x)=-2x+(1/\beta)\ln(1+\beta x)+x\ln\left[x(1+\beta x)/(1+\beta)\right]$,
while the PDF is given by Eq.~(\ref{PDFWKB}) with $x_*=1$.
The PDF variance $NS''(x_*)^{-1}=N(1+\beta)/(1+2\beta)$, indicates that such negative inhibition decreases the PDF variance by a factor of $2$ at most.

Adding negative binomial EN into the protein degradation rate, in the white noise limit, the momentum is given by Eq.~(\ref{pxgamma-main}) with $f(x)=(1+\beta)/(1+\beta x)$,
while the PDF is given by Eq.~(\ref{PDFWKB}). Using Eq.~(\ref{wnvargamma-main}) with $f(x)=(1+\beta)/(1+\beta x)$ and $x_*=1$, the observed variance in this case becomes
\begin{equation}
\sigma_{obs}^2=\frac{N(1+\beta)}{1+2\beta}(1+V\tau_c).
\end{equation}
This result indicates that negative inhibition can eliminate EN. Indeed, $\sigma_{obs}^2$ returns to its non-regulated value without EN, $N$, when the inhibition strength satisfies $\beta=V\tau_c/(1-V\tau_c)$, which holds as long as $V\tau_c<1$. That is, our choice of negative inhibition with $h=1$ can only attenuate moderate EN, and can reduce the observed variance at most by a factor of $2$.

In the adiabatic limit, we can find the PDF using Eqs.~(\ref{adiapdfgam-main}) and (\ref{fungam-main}) (see Appendix B for details), with $g(\xi)=1/(2\beta\xi)(-\xi+\sqrt{\xi^2+4\beta(\beta+1)\xi})$. Using Eq.~(\ref{adiavargam-main}) with $f(x)=(1+\beta)/(1+\beta x)$ and $f'(x_*=1)=-\beta/(\beta+1)$, the observed variance is
\begin{equation}\label{varnegreg}
\sigma_{obs}^2=\frac{N(1+\beta)}{1+2\beta}\left[1+\frac{V(1+\beta)}{1+2\beta}\right].
\end{equation}
Figure \ref{fig:NegativeRegulatedGene}(a-c) shows good agreement between theory and simulations over a wide range of parameters. Again, we see that negative inhibition can eliminate EN when $\beta=(\sqrt{1+4V}+2V-1)/[2(2-V)]$. Here, the maximum EN that can be attenuated for this particular choice of inhibition is $V=N\sigma_{ex}^2=2$.

\subsection{Higher order inhibition}
We now consider a more generic inhibition function $f(x)=(1+\beta)/(1+\beta x^h)$ with arbitrary Hill-coefficient $h$. Here, in the white-noise limit, we find
\begin{equation}
\sigma_{obs}^2=\frac{N(1+\beta)}{1+\beta(h+1)}(1+V\tau_c).
\end{equation}
In this case, EN can be eliminated by taking $\beta=V\tau_c/(h-V\tau_c)$, which holds for $V\tau_c<h$. In the adiabatic limit we obtain
\begin{equation}\label{varnegreghigh}
\sigma_{obs}^2=\frac{N(1+\beta)}{1+\beta(h+1)}\left[1+\frac{V(1+\beta)}{1+\beta(h+1)}\right].
\end{equation}
Here, EN is eliminated when $\beta=(h\sqrt{1+4V}+2V-h)/[2(h(h+1)-V)]$, which can be achieved as long as $V<V_{max}=h(h+1)$. One can see that as $h$ is increased, this inhibition mechanism becomes more efficient in eliminating EN.

\begin{figure}[h]
\begin{center}
\includegraphics{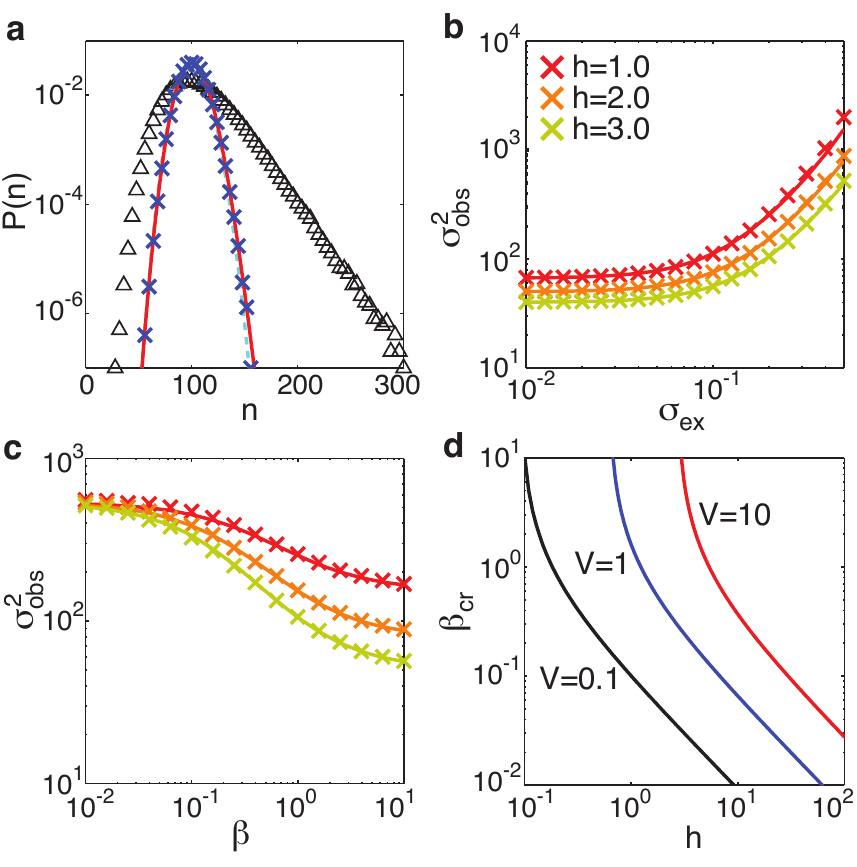}
\end{center}
\caption{(Color online) Comparison of theory (lines) and stochastic simulations (symbols) for the self-inhibited gene model with $N=100$ and adiabatic EN $\tau_c=100$. (a) Probability distributions for (solid line) theory and (x's) numerics with $\beta=\beta_{cr}=1.09$ and $h=3$. Dotted line shows the Poisson distribution for the model with only intrinsic noise and triangles show the distribution in the absence of negative inhibition. (b) Observed variance vs EN strength with $\beta=1.0$ and various values of $h$. (c) Observed variance vs inhibition strength $\beta$ for $\sigma_{ex}=0.2$ and for the same $h$ values as in (b). (d) The critical $\beta$ and $h$ values that exactly cancel EN for a given relative EN strength $V=\sigma^2_{ex}/\sigma^2_{int}$.}
\label{fig:NegativeRegulatedGene}
\end{figure}

We wanted to examine the relationship between the critical inhibition strength $\beta_{cr}$ that will exactly eliminate EN and the inhibition order. We calculated $\beta_{cr}$ across $h$ values ranging from $0.1$ to $100$ for various values of $V$. Our analytical framework provides a significant advantage over simulations for studying such large parameters spaces. Figure \ref{fig:NegativeRegulatedGene}(d) shows that cooperativity in inhibition is a necessary feature for systems that dampen strong EN.

\subsection{Self-promoting gene}

Lastly, we consider the case of a self-promoting gene. Self-promoting genes can serve as genetic switches, allowing the cell to change between two alternate expression states. The rate equation now satisfies Eq.~(\ref{RE}) where we take the production rate to be $f(x)=\alpha_0+(1-\alpha_0)\theta(x-x_0)$ --- a step function imitating a Hill-like function with a high Hill coefficient of positive feedback~\cite{Assaf2013end}. Here $\alpha_0<x_0<1$, where $\alpha_0$ is the protein concentration in the $off$ state, $x_0$ is the threshold concentration, and $N$ is the protein abundance in the $on$ state such that $x=n/N$. Unlike previously~\cite{Assaf2013end}, our derivation here is generic and does not require $x_0$ to be near one of the metastable states $\alpha_0$ or $1$.

In the absence of EN, the mean switching time (MST) from the $off$ to the $on$ states and vice versa, $\tau_{off\to on}$ and $\tau_{on\to off}$, can be calculated using the master equation~(\ref{ME}) and employing the WKB approximation (see \textit{e.g.}, Ref.~\cite{Assaf2013end}). Indeed, assuming that we start from the vicinity of the $off$ or $on$ metastable states we find the momenta $p_{off}(x)=\ln (x/\alpha_0)$ and $p_{on}(x)=\ln (x)$, where we have used the fact that $f(x)=\alpha_0$ for $x<x_0$ and $f(x)=1$ for $x>x_0$. The corresponding action functions are $S_{off}(x)=x\ln (x/\alpha_0)-x$, and $S_{on}(x)=x\ln x-x$. Therefore, the (logarithm of the) MSTs are given in the leading order of $N\gg 1$ by the accumulated action between the corresponding stable metastable state and unstable fixed point~\cite{Dykman1994lfo}
\begin{equation}\label{MSTs}
\ln\,\tau_{off\to on}\simeq N[S(x_0)-S(\alpha_0)]=N\left(\!x_0\!\ln \frac{x_0}{\alpha_0}\!-\!x_0\!+\!\alpha_0\!\right),
\end{equation}
and $\tau_{on\to off}$ coincides with $\tau_{off\to on}$ upon replacing $\alpha_0$ by $1$. For brevity, below we only present the results for the $off\to on$ switch. All the results related to the $on\to off$ switch are identical upon replacing $\alpha_0$ by $1$. Note, that in the absence of EN, $\tau_{off\to on}$ and $\tau_{on\to off}$ are comparable when $x_0=(1-\alpha_0)/\ln(1/\alpha_0)$.
Eq.~(\ref{MSTs}) can be simplified in the bifurcation limit $x_0-\alpha_0\ll \alpha_0$. Here, we find $\ln\tau_{off\to on}\simeq (N/(2\alpha_0))(x_0-\alpha_0)^2$~\cite{kamenev2008extinction,dykman2008disease}.

Now, we add negative binomial EN to the protein's degradation rate. In the white-noise limit, integrating over the momentum~(\ref{pxgamma-main}) with $f(x)=\alpha_0$ for $x<x_0$, the action function becomes
$S_{off}(x)=1/(V\tau_c)\left\{\Omega_{\alpha_0}(x)+\ln[V\tau_c x-1 + \Omega_{\alpha_0}(x)]\right.$ $\left.+V\tau_c x\left[\ln\left((x/2)(1-V\tau_c x+\Omega_{\alpha_0}(x))\right)-1\right]\right\}$,
where $\Omega_{\alpha_0}(x)=\sqrt{(1-\tau_c V x)^2+4\alpha_0 \tau_c V}$.
Therefore, the MST reads
\begin{equation}\label{MSTwhite}
\ln\,\tau_{off\to on}\simeq N[S_{off}(x_0)-S_{off}(\alpha_0)].
\end{equation}
In the bifurcation limit $x_0-\alpha_0\ll \alpha_0$, we find $\ln\tau_{off\to on}\simeq [N/(2\alpha_0)](x_0-\alpha_0)^2/(1+\alpha_0 V\tau_c)$~\cite{Assaf2013end}.

In the adiabatic limit, we need to optimize the cost of switching from one metastable state to the other given noise magnitude $\xi$ against the probability of choosing noise magnitude $\xi$~\cite{Assaf2013end}. To do so we use Eq.~(\ref{fungam-main}), where the upper integration limit is the unstable fixed point $x_0$, and for the $off\to on$ switch the lower limit is the stable fixed point given noise magnitude $\xi$, $g(\xi)=\alpha_0/\xi$.
As a result, the cost function~(\ref{fungam-main}) is a function of $\xi$ only, and reads $\Phi_{off}(\xi)=x_0\left[\ln(x_0\xi/\alpha_0)-1\right]+\alpha_0/\xi+(\xi-\ln\xi-1)/V$.
Therefore, in the leading order, the MST reads
\begin{equation}\label{switchingratesadia}
\ln\,\tau_{off\to on}\simeq N\Phi_{off}(\xi_*^{off}),
\end{equation}
where $\xi_*^{off}=\frac{1}{2}\left[1-Vx_0+\sqrt{(Vx_0-1)^2+4V\alpha_0}\right]$
is the saddle point satisfying $\Phi_{off}'(\xi_*^{off})=0$.

\begin{figure}[h]
\begin{center}
\includegraphics{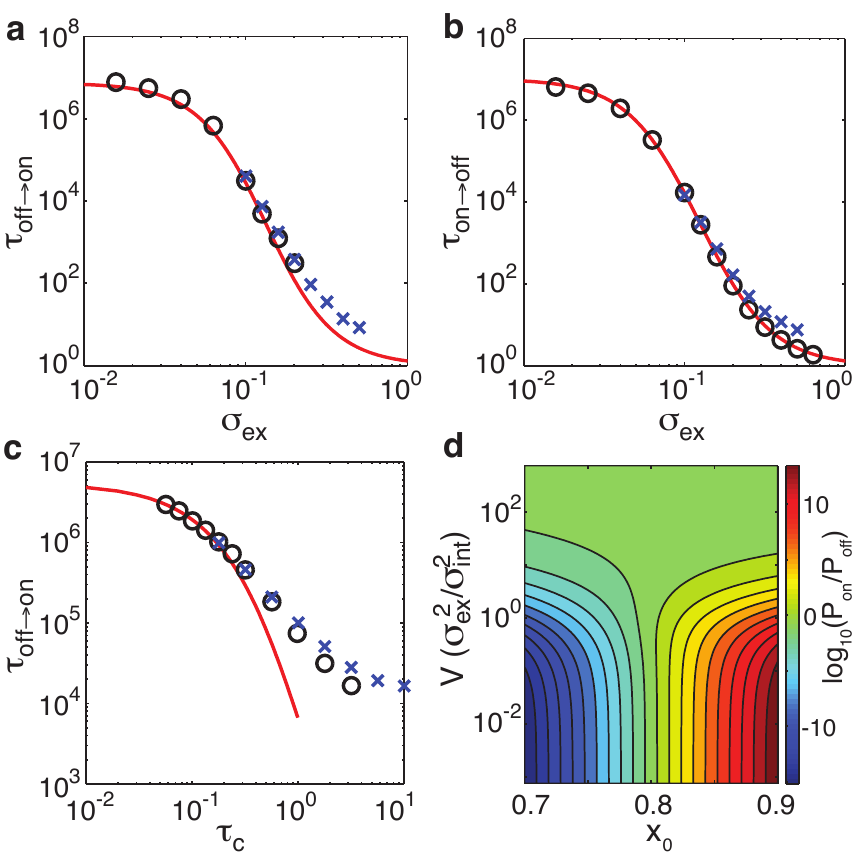}
\end{center}
\caption{(Color online) Comparison of mean switching times from analytical theory (lines) given by Eq.~(\ref{MSTwhite}), numerical solution of the Hamilton equations (o's), and stochastic simulations (x's) for the self-promoting gene model with $N=750$, $a_0=0.63$, and $x_0=0.80$. (a) The MST from the $off$ to the $on$ state vs EN strength for white noise $\tau_c=0.1$. (b) The MST from $on$ to $off$ for $\tau_c=0.1$. (c) The MST vs EN correlation time for $\sigma_{ex}=0.0365$. (d) A heat map showing the relative probability for the system to be in the $on$ vs $off$ metastable state [using Eq.~(\ref{switchingratesadia})] according to the position of the barrier $x_0$ and the relative strength $V$ of the EN. Here the EN is taken to be adiabatic.}
\label{fig:SelfRegulatedGene}
\end{figure}

Figure \ref{fig:SelfRegulatedGene}(a-b) compares our analytical theory with stochastic simulations and the numerical solution of Hamilton equations~(\ref{HEgamma-aux}) (see Methods section). Good agreement is seen in the white noise limit (similar agreement is seen in the adiabatic case), where the numerical solution of the Hamilton equations allows us to explore parameter ranges that are inaccessible by stochastic simulation due to the long MSTs. Moreover, as can be seen in Figure \ref{fig:SelfRegulatedGene}(c) the underlying Hamilton equations~(\ref{HEgamma-aux}) capture the correct dynamics well into intermediate correlation time ranges, $\tau_c={\cal O}(1)$, where the white noise approximation breaks down.

Finally, we were interested in studying the effect of EN on population dynamics. We used Eq.~(\ref{switchingratesadia}) to calculate the relative fraction of the population in the $on$ vs $off$ state as a function of both the positive feedback threshold $x_0$ and of $V$, shown in Figure \ref{fig:SelfRegulatedGene}(d). With zero or low EN the behavior of the population with regard to $x_0$ is very homogeneous. Only for a very small range of $x_0$ values is a macroscopically bistable population observed (\textit{e.g.}, at least $1$ part in $100$). If $x_0$ is not tuned very precisely, no heterogeneity is observed. As the EN increases, though, the range of macroscopic bistability increases dramatically. For $V \geq 5$, well within the range of EN observed in biological systems, the population exhibits macroscopic heterogeneity across the entire range of $x_0$ sampled. This effect is less pronounced, but still present for white EN (Figure S7, see~\cite{SuppInfo}).

\section{Conclusions}

We have presented a new formalism for studying EN in gene expression circuits, allowing us to quantify how EN affects the PDFs, variances, and MSTs in various genetic circuits. EN is likely present in multiple forms and at multiple time scales~\cite{Hensel2012sed} in a majority of processes in living cells. Understanding how EN alters the dynamics of gene networks is key for developing detailed models of genetic and regulatory processes.

Our results from studying the effect of EN on simple genetic motifs have shown that EN has a dominant influence on the system's behavior. Analyzing experimental single-cell distributions without accounting for extrinsic noise is unlikely to provide meaningful interpretation. More work needs to be done to experimentally characterize the details of the extrinsic fluctuations that cellular reactions rates experience.

Finally, EN seems to provide a distinct advantage for populations wishing to use bistability as a bet hedging strategy. Without EN, the parameters needed to have meaningful population heterogeneity in a given condition are exponentially sensitive. With EN, populations are able to explore a variety of states across a wide range of parameters.

\section{Methods}

\subsection{Monte Carlo simulations of the auxiliary circuit}
Monte Carlo simulations with negative binomial EN were performed using two auxiliary species, $a_1$ and $a_2$, to model the EN. They can be thought of in terms of a generic mRNA and protein, respectively. The dynamics of these species are given by two birth and two death processes:
\belowdisplayskip=0pt
\begin{align*}
\varnothing & \xrightarrow{v_1} a_1, & a_1 & \xrightarrow{v_2} a_1 + a_2,\\
a_1 & \xrightarrow{d_1} \varnothing, & a_2 & \xrightarrow{d_2} \varnothing,\\
\end{align*}
Here, $\varnothing$ is a symbol for the empty set, i.e., a species is created from nothing or destroyed into nothing. The rates are given by:
\begin{align*}
v_1 & = \frac{1}{\tau_c}\frac{K}{K \sigma^2_{ex} - 1}, & v_2 & = \frac{\omega (K \sigma^2_{ex} -1)}{\tau_c}\\
d_1 & = \frac{\omega}{\tau_c}, & d_2 & = \frac{1}{\tau_c}.\\
\end{align*}
$K$ is the mean copy number of species $a_2$, $\sigma^2_{ex}$ is the desired EN strength, and $\tau_c$ is the desired EN correlation time. To remind the reader, the negative binomial parameters, $\alpha$ and $\beta$, defining the distribution~(\ref{NBdist}) are related to $v_1$ and $v_2$ in the following manner: $v_1=\alpha/\tau_c$, and $v_2=\omega\beta/\tau_c$, where $\alpha=K/(K\sigma_{ex}^2-1)$ and $\beta=K\sigma_{ex}^2-1$. In all simulations $\omega$ was set to $100$. To avoid negative rates, one must have $K \sigma^2_{ex} > 1$. Therefore, the value used for $K$ limits the lower bound of the EN that can be simulated using a particular set of parameters. In the limit as $K \sigma^2_{ex} \rightarrow 1$ the variance approaches the Poissonian variance of a birth death process centered on $K$. Larger $K$ allows for smaller EN to be simulated. Outside of this limitation, $K$ has no influence on the EN properties. It does, however, influence the computational efficiency of the simulation. Both larger and smaller values of $K$ increase the runtime of the simulations. We used a value of $K=20,000$ throughout this work, which leads to a lower bound for the EN studied of $\sigma_{ex} > 0.007$ and provides reasonable runtimes.

Figure S1, see~\cite{SuppInfo}, shows that the mean and variance of the auxiliary species are as expected during the simulations. Figure S2, see~\cite{SuppInfo}, shows that the auxiliary species has the expected autocorrelation time. All simulations were performed using the standard Gillespie algorithm \cite{Gillespie1977ess} using the Lattice Microbes software \cite{Roberts2013lmh}.

The auxiliary species $a_2$ was coupled to a reaction to be fluctuated by including it as an additional species participating in the reaction and adjusting the reaction rate constant such that the mean equals the original value. For example, to model a fluctuating birth rate for protein $n$ with mean copy number $N$, the reaction $\varnothing \xrightarrow{N} n$ becomes  $a_2  \xrightarrow{N/K} n + a_2$. Similarly, to model a fluctuating death rate $n \xrightarrow{1} \varnothing$ becomes  $n + a_2  \xrightarrow{1/K} a_2$. In the above equations, $a_2$ appears on both sides to indicate that it is neither created nor destroyed by the reaction.

\subsection{Numerical solutions of the Hamilton equations}
In this section we use the shooting method~\cite{kamenev2008extinction,Roma2005ope,dykman2008disease} to find a numerical solution to the set of Hamilton equations~(\ref{HEgamma-aux}) in the case of arbitrary correlation time $\tau_c$. We focus on the case of the self-promoting gene, which gives rise to switching between metastable phenotypic states. Here there are three fixed points in the language of the deterministic rate equations, two stable points at $x_{off}=\alpha_0$ and $x_{on}=1$ and one unstable point at $x_s=x_0$. We are interested to numerically compute the trajectories, $\mathbf{z}_{on}(t)$ and $\mathbf{z}_{off}(t)$, corresponding to the optimal paths along which switching from the $on \to off$ and $off \to on$ occurs, respectively. Below we consider the trajectory $\mathbf{z}_{on}(t)$, where the analysis of $\mathbf{z}_{off}(t)$ is similar.

Let us denote by $t_i=0$ and $t_f$ the initial and final simulation times, respectively. The initial condition is given by $\mathbf{z}_{on}(0)=\mathbf{x}_{on}+\delta \mathbf{v}$ where $\mathbf{x}_{on}=(x=x_{on}, p_x=0, \xi=1, p_{\xi}=0)$ is the corresponding fixed point in the 4D phase space and $\mathbf{v}$ is the initial direction of the trajectory, see below. Here $\delta$ is chosen to be small, but not too small to balance between simulation runtime and accuracy.  The final condition is that the trajectory reaches the close vicinity of $\mathbf{x}_s=(x_s,0,1,0)$, namely that $|\mathbf{z}_{on}(t_f)-\mathbf{x}_s|\ll 1$. [From there, the assumption is that the system flows almost deterministically to $\mathbf{x}_{off}=(\alpha_0,0,1,0).$]

In order to find the initial direction of the trajectory we linearize the Hamilton equations~(\ref{HEgamma-aux}) in the vicinity of $\mathbf{x}_{on}$. This allows us to find the eigenvalues and eigenvectors in the vicinity of $\mathbf{x}_{on}$. Since the switching trajectory leaves the  fixed point $\mathbf{x}_{on}$ along its unstable manifold, we are only considering the eigenvectors $\mathbf{v}_1$ and $\mathbf{v}_2$ that correspond to the two positive eigenvalues $\lambda_1$ and $\lambda_2$. (Note that the two other eigenvalues satisfy $\lambda_3=-\lambda_1$ and $\lambda_4=-\lambda_2$ such that $\sum_i \lambda_i=0$.) As a result, we take the initial direction to be $\mathbf{v}=\mathbf{v}_1\cos(\alpha)+\mathbf{v}_2\sin(\alpha)$, where it is assumed that each of the eigenvectors is normalized to unity, and $0\leq\alpha\leq 2\pi$. Finally, we search over all possible values of $\alpha$ until we find the best choice that satisfies the above initial and final conditions. Note that along $\mathbf{z}_{on}(t)$, the initial conditions $\partial_t x(0)<0$ and $\partial_t p_x(0)<0$ are satisfied.

This search over $\alpha$ is optimized by performing a binary search. Each time an initial condition is chosen, the set of equations is solved numerically by using a Matlab numerical solver, and we compare the final condition to $\mathbf{x}_s$. The search is terminated when we have sufficiently converged to the final condition.  After successfully determining the trajectory we perform a numerical integration in order to find the accumulated action. We do so by using the formula $\Delta S=\int_0^{t_f} [p_x(t)\dot{x}+p_{\xi}(t)\dot{\xi}]dt\simeq\sum_{i=0}^{N-1}[x(t_{i+1})-x(t_i)]\times[p_x(t_{i+1})+p_x(t_i)]/2+[\xi(t_{i+1})-\xi(t_i)]\times[p_{\xi}(t_{i+1})+p_{\xi}(t_i)]/2$, where $t_0=0$ and $t_N=t_f$. This result gives us the logarithm of the mean switching time divided by $N$. An example can be seen in Figure S8 in~\cite{SuppInfo}.

\begin{acknowledgments}
We thank Naftali R. Smith for useful discussions. This work was supported by Grant No. 300/14 of the Israel Science Foundation.
\end{acknowledgments}

\appendix


\setcounter{equation}{0}
\section{mRNA-protein auxiliary circuit}
In this section, we  derive the stationary PDF of the auxiliary protein. This PDF determines the extrinsic noise (EN) statistics of the degradation rate of the protein of interest. The choice of negative binomial statistics used in the main text for the reaction rate seems quite natural. Indeed, in genetic circuits of a non-regulated gene, if the mRNA is short lived, the proteins' stationary PDF is given by a negative binomial distribution~\cite{Paulsson2000rsf,Cai2006spe}. As a result, if this auxiliary protein affects the degradation rate of our protein of interest, this rate will fluctuate with negative binomial statistics.

In the auxiliary mRNA-protein circuit, mRNAs are transcribed at a rate $\alpha/\tau_c$ and degrade with rate $\omega/\tau_c$, while proteins are translated at a rate $\omega \beta/\tau_c$ and degrade at a rate $1/\tau_c$, which insures that the correlation time of the auxiliary proteins is $\tau_c$. We assume a short-lived mRNA such that $\omega\gg 1$. The master equation describing the probability to find $m$ mRNAs and $k$ proteins satisfies:
\begin{eqnarray}\label{auxmaster}
&&\dot{P}_{m,k}=\frac{\alpha}{\tau_c}(P_{m-1,k}\!-\!P_{m,k})+\frac{\omega}{\tau_c}[(m+1)P_{m+1,k}\!-\!mP_{m,k}]\nonumber\\
&&+\frac{\omega \beta m}{\tau_c}(P_{m,k-1}\!-\!P_{m,k})+\frac{1}{\tau_c}[(k+1)P_{m,k+1}\!-\!kP_{m,k}]\!.
\end{eqnarray}
We denote the auxiliary mRNA and protein concentrations by $z=m/K$ and $\xi=k/K$, respectively, where $K=\alpha\beta$ is the auxiliary protein's abundance. We now use a dissipative version of the WKB approximation, see \textit{e.g.}, Refs.~\cite{Dykman1994lfo,Bender1999,kessler2007extinction,escudero2009switching}. Employing the WKB ansatz $P_{m,k}=P(z,\xi)\sim e^{-KS(z,\xi)}$, we arrive at a Hamilton-Jacobi equation $H=0$ with Hamiltonian
\begin{eqnarray}\label{Hamaux}
&&H(z,p_z,\xi,p_{\xi})=\frac{\alpha}{K}(e^{p_z}-1)+\omega z(e^{-p_z}-1)\nonumber\\
&&+\beta\omega z(e^{p_{\xi}}-1)+\xi(e^{-p_{\xi}}-1),
\end{eqnarray}
where $p_z=\partial_z S(z,\xi)$ and $p_{\xi}=\partial_{\xi} S(z,\xi)$ are the associated mRNA and protein momenta.
This yields the following Hamilton equations
\begin{eqnarray}\label{HE2D}
\dot{z}&=&\frac{\alpha}{K} e^{p_z}-\omega ze^{-p_z},\nonumber\\
\dot{p}_z&=&-\omega(e^{-p_z}-1)-\beta\omega(e^{p_{\xi}}-1),\nonumber\\
\dot{\xi}&=&\beta\omega z e^{p_{\xi}}-\xi e^{-p_{\xi}},\nonumber\\
\dot{p}_{\xi}&=&1-e^{-p_{\xi}}.
\end{eqnarray}
For $\omega\gg 1$, the mRNA lifetime is short compared to that of the protein. In this case, $z$ and $p_z$ equilibrate much faster than $\xi$ and $p_{\xi}$, and we can adiabatically eliminate the mRNA species~\cite{assaf2008noise,Shahrezaei2008ads}. As a result, putting $\dot{z}=\dot{p}_z=0$ we find $z=z(\xi,p_{\xi})$ and $p_z=p_z(\xi,p_{\xi})$. Plugging this into the Hamiltonian~(\ref{Hamaux}) we arrive at the reduced Hamiltonian for the auxiliary protein only~\cite{Vardi2013bye}
\begin{equation}\label{Hredaux}
H_{r}(\xi,p_{\xi})=\frac{1}{\beta}\left[\frac{1}{1+\beta(1-e^{p_{\xi}})}-1\right]+\xi(e^{-p_{\xi}}-1),
\end{equation}
which effectively includes mRNA fluctuations. Solving the Hamilton-Jacobi equation $H_r(\xi,p_{\xi})=0$ we find
\begin{equation}\label{protmom}
p_{\xi}=\ln[(1+\beta)\xi/(1+\beta\xi)],
\end{equation}
and thus, the action becomes
\begin{equation}\label{actionmRNA}
S(\xi)=\xi\ln\left[\frac{(1+\beta)\xi}{1+\beta \xi}\right]-\frac{1}{\beta}\ln(1+\xi\beta).
\end{equation}
As a result, the stationary PDF to find $k$ copies of the auxiliary protein is given by $P(k)\sim e^{-K[S(k/K)-S(1)]}$ [see Eq.~(\ref{PDFWKB}) in the main text], where $K=\alpha\beta$ is the protein abundance. This distribution, when properly normalized, coincides at $k\gg 1$ with the negative binomial distribution
\begin{equation}\label{NBdist}
P_k=\frac{\Gamma(\alpha+k)}{\Gamma(k+1)\Gamma(\alpha)}\left(\frac{\beta}{\beta+1}\right)^k\,\left(\frac{1}{\beta+1}\right)^{\alpha}.
\end{equation}

Interestingly, these results can give us insight on the mRNA fluctuations that are implicitly incorporated in the protein-only model [Eq.~(\ref{Hredaux})], after eliminating the fast mRNA variable. Indeed, by comparing Eq.~(\ref{protmom}) with the momentum in the protein-only model [Eq.~(\ref{momentum})], we find that  mRNA fluctuations emanating from this unregulated mRNA-protein circuit can be effectively accounted for by taking a protein-only model with a \textit{modified} production rate $f(x)=(1+\beta x)/(1+\beta)$. This production rate becomes $1$ in the limit of small burst size $\beta\to 0$, but becomes $\xi+(1-\xi)/\beta$ in the limit of large burst size $\beta\gg 1$, which yields a much wider distribution with variance $N\beta\gg N$. Importantly, this modified production rate gives rise to a protein PDF that coincides with the negative binomial distribution at $k\gg 1$.

\setcounter{equation}{0}
\section{Analysis of the Hamiltonian combining IN and EN}
In this section we will derive the stationary PDF of the proteins of interest for generic production rate $f(x)$, where the degradation rate fluctuates due to EN with negative binomial statistics and correlation time $\tau_c$. In order to do so, we will analyze the Hamilton equations emanating from Hamiltonian~(\ref{2D-ham}) in the main text. In particular, we will find approximate solutions for the protein PDF in the limits of short- and long-correlated EN.

Using Hamiltonian~(\ref{2D-ham}) the corresponding Hamilton equations read
\begin{eqnarray}\label{HEgamma-aux}
\dot{x}&=&f(x)e^{p_x}-\frac{x\tilde{\xi}}{\rho} e^{-p_x},\nonumber\\
\dot{p}_x&=&-f'(x)(e^{p_x}-1)-\frac{\tilde{\xi}}{\rho}(e^{-p_x}-1)\nonumber\\
\dot{\tilde{\xi}}&=&\frac{\rho e^{p_{\tilde{\xi}}}}{\tau_c[1+\beta(1-e^{p_{\tilde{\xi}}})]^2}-\frac{e^{-p_{\tilde{\xi}}}\tilde{\xi}}{\tau_c},\nonumber\\
\dot{p}_{\tilde{\xi}}&=&-\frac{(e^{-p_{\tilde{\xi}}}-1)}{\tau_c}-\frac{x}{\rho}(e^{-p_x}-1).
\end{eqnarray}
To remind the reader, $\rho=K/N$ is the abundances ratio of the auxiliary protein and protein of interest and $\tilde{\xi}=\rho\xi=k/N$ is a rescaled noise variable, while $\alpha=1/(\sigma_{ex}^2-1/K)$ and $\beta=K\sigma_{ex}^2-1$. Hamilton equations~(\ref{HEgamma-aux}) can be solved numerically for any value of $\tau_c$, see Methods section. This numerical solution provides the statistics of interest in the leading order, and is far more efficient than performing numerical Monte-Carlo simulations, especially for short-correlated EN, see main text. Importantly, the cases of fast and slow dynamics of the EN can be studied analytically, see below.

\subsection{White-noise limit of EN}
In the white-noise limit, $\tau_c\ll 1$, the dynamics of the auxiliary protein is fast. As a result, $\tilde{\xi}(t)$ and $p_{\tilde{\xi}}(t)$ equilibrate fast compared to $x$ and $p_x$, and we can look for slowly-varying $x$ and $p_x$ dependent solutions of the third and fourth Hamilton equations~(\ref{HEgamma-aux}). This yields in the leading order of $\tau_c\ll 1$
\begin{equation}\label{xieffgamma}
\xi^{eff}=1-2(1-e^{-p_x})V x \tau_c,
\end{equation}
where we have defined $V\equiv N\sigma_{ex}^2$ as the ratio between the relative EN and IN variances. Note, that the value of $p_{\tilde{\xi}}^{eff}={\cal O}(\tau_c)\ll 1$ does not enter the equations for $\dot{x}$ and $\dot{p}_x$. As expected, Eq.~(\ref{xieffgamma}) as well as the results below are independent of the arbitrary choice of the auxiliary protein abundance $K$. Plugging $\xi^{eff}(x,p_x)$ from Eq.~(\ref{xieffgamma}) into the first and second Hamilton equations~(\ref{HEgamma-aux}) we arrive at an effective 1D white-noise Hamiltonian~\cite{Kamenev2008hce,Assaf2013end,Assaf2013cdf}
\begin{equation}\label{HredNoiseGamma}
H(x,p_x)=f(x)(e^{p_x}-1)+x(e^{-p_x}-1)+x^2(e^{-p_x}-1)^2V\tau_c.
\end{equation}
Solving the Hamilton-Jacobi equation $H=0$ we find the momentum
\begin{equation}\label{pxgamma}
p_x=\ln\left\{\frac{x}{2f(x)}\left[1\!-\!V\tau_c x\!+\!\sqrt{(V\tau_c x\!-\!1)^2+4V f(x)\tau_c}\right]\right\}.
\end{equation}
The PDF can be formally found by integrating Eq.~(\ref{pxgamma}) to find the corresponding action $S(x)=\int^x p_x(x')dx',$ and by using Eq.~(\ref{PDFWKB}).

\subsection{Adiabatic limit of EN}
In the adiabatic limit, $\tau_c\gg 1$, we can assume the EN is almost stationary. As a result, the stationary PDF of the proteins satisfies~\cite{Kamenev2008hce,Assaf2013end,Assaf2013cdf}
\begin{equation}\label{pdfA}
P_n=\int_{-\infty}^{\infty}P(\xi)P(n|\xi)d\xi,
\end{equation}
where $P(n|\xi)$ is the probability to find $n$ proteins given noise magnitude $\xi$, and $P(\xi)$ is the probability to find EN magnitude $\xi$.
For simplicity we will take the EN to be gamma distributed, $P(\xi)=\tilde{\beta}^{-\alpha}/\Gamma(\alpha)\,\xi^{\alpha-1}e^{-\xi/\tilde{\beta}}$. Here $\alpha=1/\sigma_{ex}^2$ and $\tilde{\beta}=\beta/K\simeq \sigma_{ex}^2$, which guarantees that the mean is $1$ and the variance is $\sigma_{ex}^2$. As can be checked, the gamma distribution becomes a good approximation of the negative binomial distribution when $K$ is sufficiently large.

With these values of $\alpha$ and $\tilde{\beta}$, we find
\begin{equation}
P(\xi)\simeq \frac{1}{\xi\sqrt{2\pi\sigma_{ex}^2}}e^{(1/\sigma_{ex}^2)(1+\ln \xi-\xi)},
\end{equation}
which holds as long as $\sigma_{ex}<1$. As a result, the PDF to find $n$ proteins [Eq.~(\ref{pdfA})] becomes
\begin{equation}\label{pdfB}
P_n=\int_{-\infty}^{\infty}\frac{1}{\xi\sqrt{2\pi\sigma_{ex}^2}}P(n|\xi)e^{-(\xi-\ln \xi-1)/\sigma_{ex}^2}d\xi,
\end{equation}
where $$P(n|\xi)=Ae^{-N\int_{g(\xi)}^x\ln\frac{y\xi}{f(y)}dy},$$
and $A=A(\xi)$ is a normalization constant. Here, we have used the fact that given $\xi$, the momentum along the optimal path (zero-energy Hamiltonian) satisfies $p_x(x,\xi)=\ln [x\xi/f(x)]$, and the fixed point given noise magnitude $\xi$ satisfies the equation $f(x)=x\xi$ and is given by $x(\xi)=g(\xi)$.

To proceed, we rewrite the integral in Eq.~(\ref{pdfB}) as
\begin{equation}\label{Pnadia}
P_n=\int_{-\infty}^{\infty}\frac{B}{\xi}e^{-N\Phi(x,\xi)}d\xi,
\end{equation}
where $B$ contains the preexponential factors including all normalization constants and
\begin{eqnarray}\label{fungam}
\Phi(x,\xi)=\int_{g(\xi)}^x\ln\frac{y\xi}{f(y)}dy+\frac{\xi-\ln\xi-1}{V},
\end{eqnarray}
is the cost function that we need to optimize, see main text. Now, we use the fact that $N\gg 1$ and employ the saddle-point approximation. The saddle point is obtained at $\partial_{\xi}\Phi(x,\xi)=0$, which yields the following algebraic equation
\begin{equation}\label{dergam}
\frac{\partial \Phi}{\partial\xi}=\frac{x-g(\xi)}{\xi}+\frac{1}{V}-\frac{1}{V\xi}=0,
\end{equation}
where we have used the Leibniz integral rule when differentiating Eq.~(\ref{fungam}), and $g(\xi)$ is defined above. Solving the equation $V[x-g(\xi)]+\xi-1=0$ for $\xi$ yields the optimal noise magnitude $\xi_*(x)$. Plugging $\xi_*(x)$ into $\Phi(x,\xi)$  we find the PDF in the adiabatic limit, which is given by Eq.~(\ref{adiapdfgam-main}) in the main text, where  $\partial_{\xi\xi}\Phi(x,\xi)=[1-Vg'(\xi)]/(V\xi)$.

The variance of this PDF can be explicitly calculated. It is given by the second derivative of $\Phi[x,\xi=\xi_*(x)]$ [Eq.~(\ref{fungam}) when plugging $\xi=\xi_*(x)$] with respect to $x$, evaluated at $x=x_*$
\begin{equation}\label{secdergam}
N var^{-1}=\left.\frac{d^2 \Phi[x,\xi=\xi_*(x)]}{d x^2}\right|_{x=x_*},
\end{equation}
where $x_*$ is the unperturbed fixed point (with $\xi=1$) satisfying $x_*=f(x_*)$. To carry out this calculation analytically we need to solve Eq.~(\ref{dergam}) and find the optimal noise magnitude $\xi_*$. We recall that the variance is calculated in the vicinity of the unperturbed fixed point $x\simeq x_*$. Let us assume a-priori that $|\xi-1|\ll 1$ in the vicinity of $x\simeq x_*$. Then, we can expand $g(\xi)$ in the vicinity of $\xi=1$, $g(\xi)\simeq g(1)+g'(1)(\xi-1)$. However, since $g(1)$ is the solution of the equation $x\xi=f(x)$ at $\xi=1$, we have $g(1)=x_*$. Therefore, we have
\begin{equation}\label{gxi}
g(\xi)\simeq x_*+g'(1)(\xi-1).
\end{equation}
Plugging this into Eq.~(\ref{dergam}) we find
\begin{equation}\label{xigam}
\xi_*(x)\simeq 1+\frac{V(x-x_*)}{Vg'(1)-1}.
\end{equation}
This verifies our assumption that $|1-\xi_*(x)|\ll 1$ as long as $x$ is in the close vicinity of $x_*$. Now, using Eqs.~(\ref{gxi}) and (\ref{xigam}) in Eq.~(\ref{secdergam}), performing the differentiation, and evaluating the result at $x=x_*$, we find the observed variance to be
\begin{equation}\label{adiavariancegam}
\sigma_{obs}^2=\frac{Nx_*[Vg'(1)-1]^2}{[f'(x_*)-1][2Vg'(1)-1]-V x_*},
\end{equation}
where we have used the fact that $x_*=f(x_*)$. This expression can be further simplified if we recall that $g(\xi)$ satisfies $f[g(\xi)]/g(\xi)=\xi$. Differentiating this with respect to $\xi$, evaluating the result at $\xi=1$, and using the fact that $g(1)=x_*$, we obtain $g'(1)=x_*/[f'(x_*)-1]$. Plugging this into Eq.~(\ref{adiavariancegam}) we arrive at the final result
\begin{equation}\label{adiavargam}
\sigma_{obs}^2=\frac{Nx_*}{1-f'(x_*)}\left[1+\frac{Vx_*}{1-f'(x_*)}\right].
\end{equation}
Note, that throughout these calculations we have assumed that the mean of the PDF remains at $x=x_*$, and calculated the variance accordingly. This assumption is accurate as long as the EN magnitude is not too strong, $\sigma_{ex}^2\ll 1$, see~Figure S3(c) and Figure S4 in~\cite{SuppInfo}, which is within the range of EN observed in biological systems.

Yet, for very strong EN, the mean of the PDF shifts to the right, due to the nonlinear dependence of the fixed point on the degradation rate, and due to the corresponding slowly-decreasing right tail of the protein PDF. We will now show this explicitly in the case of the unregulated gene, for which $f(x)=1$. Let us assume $\sigma_{ex}={\cal O}(1)$ such that $V=N\sigma_{ex}^2={\cal O}(N)\gg 1$. In this strong-EN regime, the PDF is approximately given by
\begin{equation}\label{pd}
P(n)\simeq C\sqrt{\frac{N}{n}}e^{\frac{N}{V}\left(1-\frac{N}{n}-\ln\frac{n}{N}\right)},
\end{equation}
where we have used Eqs.~(\ref{adiapdfgam-main}) and (\ref{fungam-main}) in the main text, with $g(\xi)=1/\xi$ and $\xi_*(x)\simeq 1/x$, and $C=(2\pi N V)^{-1/2}$.

In order to calculate the mean of this PDF we use the equality $\langle n\rangle=\sum_n n P_n$. Doing so, and using the saddle point approximation, we find
\begin{equation}
\langle n\rangle \simeq N(1+3\sigma_{ex}^2/2).
\end{equation}
This result for the PDF mean in the case of EN in the degradation rate agrees well with simulations, see Figure S3 and Figure S4 in~\cite{SuppInfo}.

\setcounter{equation}{0}
\section{The case of EN in the production rate}
In this section we consider EN in the production rate rather than in the degradation rate. We show that while the resulting protein PDF in this case differs from the case of EN in the degradation rate, the variance of the PDF coincides in the two cases, in both the white- and adiabatic-EN limits.

We again consider EN with a negative binomial statistics and correlation time $\tau_c$. Our starting point is the 2D Hamiltonian which encodes the stochastic dynamics of the protein of interest under the influence of EN. Here, instead of EN in the degradation rate we have EN in the production rate in the form $f(x)\to\xi f(x)$, where $\xi$ satisfies $\langle\xi\rangle=1$, and fluctuates with negative binomial statistics. As a result, the Hamiltonian~(\ref{2D-ham}) in the case of EN in the degradation rate, gives way to
\begin{eqnarray}\label{2D-ham-prod}
&&H(x,p_x,\xi,p_{\xi})=\frac{\tilde{\xi}}{\rho}f(x)(e^{p_x}-1)+x(e^{-p_x}-1)\nonumber\\
&&+\frac{\rho}{\beta\tau_c}\left[\frac{1}{1+\beta(1-e^{p_{\tilde{\xi}}})}-1\right]+\frac{\tilde{\xi}}{\tau_c}(e^{-p_{\tilde{\xi}}}-1),
\end{eqnarray}
where $\tilde{\xi}$ and $\rho$ are defined above.
At this point, we can repeat the calculations done above for EN in the degradation rate. In the white noise limit we find the momentum to be
\begin{equation}\label{pxgamma-prod}
p_x=\ln\left[\frac{V\tau_c f(x)-1+\sqrt{(V\tau_c f(x)-1)^2+4V\tau_c x}}{2V\tau_c f(x)}\right],
\end{equation}
from which the PDF can be calculated via Eq.~(\ref{PDFWKB}), with $S(x)=\int^x p(x')dx'$. This PDF does not coincide with the case of EN in the degradation rate [compare Eq.~(\ref{pxgamma-prod}) with Eq.~(\ref{pxgamma})], but for weak and moderate EN, the PDFs are indistinguishable, see~Figure S3(a) in~\cite{SuppInfo}. Differentiating the momentum with respect to $x$ we find the observed variance to be
\begin{equation}\label{var-gamma-prod}
\sigma_{obs}^2=NS''(x_*)^{-1}=\frac{Nx_*(1+x_*V\tau_c)}{1-f'(x_*)},
\end{equation}
which coincides with the variance when EN is in the degradation rate.

In the adiabatic case, we again need to calculate the integral
\begin{equation}\label{Pnadia-prod}
P_n\sim\int_{-\infty}^{\infty}\frac{B}{\xi}e^{-N\Phi(x,\xi)}d\xi,
\end{equation}
where $B$ contains the preexponential factors including all normalization constants. In this case, the cost function $\Phi(x,\xi)$ takes the form
\begin{eqnarray}\label{fungam-prod}
\Phi(x,\xi)=\int_{g(\xi)}^x\ln\frac{y}{\xi f(y)}dy+\frac{\xi-\ln\xi-1}{V}.
\end{eqnarray}
Note, that the only difference between this equation and Eq.~(\ref{fungam}) is that here $\xi$ is in the denominator of the $\ln$ function, instead of the numerator. In addition, in this case $x=g(\xi)$ solves the equation $\xi f(x)=x$. Now, we use the fact that $N\gg 1$ and solve the integral~(\ref{Pnadia-prod}) via the saddle-point approximation. The saddle point is obtained at $\partial_{\xi}\Phi(x,\xi)=0$, which yields the following algebraic equation
\begin{equation}\label{dergam-prod}
\frac{\partial \Phi}{\partial\xi}=\frac{g(\xi)-x}{\xi}+\frac{1}{V}-\frac{1}{V\xi}=0.
\end{equation}
Solving the equation $V[g(\xi)-x]+\xi-1=0$ for $\xi$ we find $\xi_*(x)$, which allows finding the PDF according to Eq.~(\ref{adiapdfgam-main}), see~Figure S3(b) in~\cite{SuppInfo}. This figure emphasizes the lack of coincidence between the PDFs in the cases of adiabatic EN in the production and degradation rates.

The variance of this PDF is given by Eq.~(\ref{secdergam}), where $x=x_*$ is the unperturbed fixed point satisfying $x_*=f(x_*)$.
To carry out this calculation analytically we need to solve Eq.~(\ref{dergam-prod}) and find the optimal noise magnitude $\xi_*$. We recall that the variance is calculated in the vicinity of the fixed point $x\simeq x_*$. Assuming a-priori that $|\xi-1|\ll 1$ in the vicinity of $x\simeq x_*$, we take Eq.~(\ref{dergam-prod}) and expand $g(\xi)$ to first order in $\xi$ around $\xi=1$. By doing so, and using the fact that $g(1)=x_*$, we have $g(\xi)\simeq x_*+g'(1)(\xi-1)$, which yields
\begin{equation}\label{xigam-prod}
\xi_*(x)\simeq 1+\frac{V(x-x_*)}{Vg'(1)+1}.
\end{equation}
Indeed, we find that $|1-\xi_*(x)|\ll 1$ as long as $x$ is in the close vicinity of $x_*$. Now, we plug $\Phi(x,\xi)$ from Eq.~(\ref{fungam-prod}) and $\xi=\xi_*(x)$ from Eq.~(\ref{xigam-prod}) into Eq.~(\ref{secdergam}). Performing the differentiation twice with respect to $x$, plugging $x=x_*=f(x_*)$, using the fact that $g(\xi)$ satisfies $f[g(\xi)]/g(\xi)=1/\xi$, and evaluating the result at $\xi=1$ which yields $g'(1)=x_*/[1-f'(x_*)]$, we find the observed variance to be
\begin{equation}\label{adiavargam-prod}
\sigma_{obs}^2=\frac{Nx_*}{1-f'(x_*)}\left[1+\frac{Vx_*}{1-f'(x_*)}\right].
\end{equation}
This result coincides with the variance in the case of EN in the degradation rate, see~Figure S3(d) in~\cite{SuppInfo}.

\setcounter{equation}{0}
\section{The case of Ornstein-Uhlenbeck EN}
In this section we consider EN with different statistics. We take Ornstein-Uhlenbeck (OU) extrinsic noise with mean $\langle \xi(t)\rangle=1$ and  variance $\langle \xi(t) \xi(t') \rangle =\sigma_{ex}^2 e^{-|t-t'|/\tau_c}$ with correlation time $\tau_c>0$. Note that in our previous work~\cite{Assaf2013end} on the self-regulating-gene model, we have already used the OU noise when modeling EN. Here we develop a different and more generic formalism allowing to go beyond the bifurcation limit done previously, and to treat EN of arbitrary strength. Notably, EN with such statistics can give rise to zero or even negative reaction rates for sufficiently strong EN, which can cause, \textit{e.g.}, the divergence of the mean~\cite{bostani2012noise}. As a result, in our derivation below we implicitly assume that the noise statistics has a cutoff such that the reaction rates are always positive real numbers.

The OU process satisfies the following Langevin equation
\begin{equation}\label{lannoise}
\dot{\xi}=-(\xi-1)/\tau_c+\sqrt{2\sigma_{ex}^2/\tau_c}\;\eta(t),
\end{equation}
where $\eta(t)$ is white noise $\langle\eta(t) \eta(t')\rangle= \delta(t-t')$. Here $\eta(t)$ can be defined as the $dt\to 0$ limit of the temporally uncorrelated normal
random variable with mean $0$ and variance $1/dt$. The stationary statistics of this noise is $P(\xi)=1/\sqrt{2\pi\sigma_{ex}^2}e^{-(\xi-1)^2/(2\sigma_{ex}^2)}$.

In order to go beyond the bifurcation limit~\cite{Assaf2013end}, we are interested to describe the OU process via a discrete birth death process describable by a master equation. Defining $k\equiv K\xi$ as the noise ``copy number" in the OU process where $K\gg 1$ is an arbitrary large number, the master equation describing the probability $P_k$ to find EN copy number $k$ satisfies:
\begin{equation}\label{OUmaster}
\dot{P}_k=\lambda_{k-1}P_{k-1}+\nu_{k+1}P_{k+1}-(\lambda_k+\nu_k)P_k.
\end{equation}
Here $\lambda_k=1/(2\tau_c)(2K^2\sigma_{ex}^2-k+K)$ and $\nu_k=1/(2\tau_c)(2K^2\sigma_{ex}^2+k-K)$ are the birth and death rates, respectively. One can check that using these birth and death rates one recovers the Langevin equation for the EN ``copy number": $\dot{k}=-(k-K)/\tau_c+\sqrt{2K^2\sigma_{ex}^2/\tau_c}\;\eta(t)$, which corresponds to Eq.~(\ref{lannoise}) with $\xi=k/K$.

To study the interplay between IN and EN, we combine the EN dynamics [Eq.~(\ref{OUmaster})] with the underlying IN dynamics [Eq.~(\ref{ME})].  This yields a 2D master equation for the probability $P(n,k,t)$ to find protein copy number $n$ and noise copy number $k$, at time $t$. Similarly as in the case of negative binomial EN, using the WKB ansatz for the stationary PDF, $P_{n,k}\sim e^{-NS(n/N,k/N)}$, we arrive at a Hamilton-Jacobi equation $H=0$ with a Hamiltonian
\begin{eqnarray}\label{HamOU}
&&H^{(OU)}(x,p_x,\tilde{\xi},p_{\tilde{\xi}})=f(x)(e^{p_x}-1)+\frac{x\tilde{\xi}}{\rho}(e^{-p_x}-1)\\
&&+\frac{(2\rho^2 V-\tilde{\xi}+\rho)}{2\tau_c}(e^{p_{\tilde{\xi}}}-1)+\frac{(2\rho^2 V+\tilde{\xi}-\rho)}{2\tau_c}(e^{-p_{\tilde{\xi}}}-1),\nonumber
\end{eqnarray}
where as before $V=N\sigma_{ex}^2$, $\tilde{\xi}=\rho\xi$ and $\rho=K/N$, while $p_x=\partial_x S$, and $p_{\tilde{\xi}}=\partial_{\tilde{\xi}} S$ are the associated momenta.
This Hamiltonian encodes the stochastic dynamics of the protein of interest when its degradation rate fluctuates with OU noise.

In order to proceed, we can write down the corresponding Hamilton equations
\begin{eqnarray}\label{HE}
\dot{x}&=&f(x)e^{p_x}-\frac{x\tilde{\xi}}{\rho} e^{-p_x}\nonumber\\
\dot{p}_x&=&-f'(x)(e^{p_x}-1)-\frac{\tilde{\xi}}{\rho}(e^{-p_x}-1)\nonumber\\
\dot{\tilde{\xi}}&=&\frac{e^{p_{\tilde{\xi}}}}{2\tau_c}(2\rho^2 V-\tilde{\xi}+\rho)-\frac{e^{-p_{\tilde{\xi}}}}{2\tau_c}(2\rho^2 V+\tilde{\xi}-\rho)\nonumber\\
\dot{p}_{\tilde{\xi}}&=&\frac{1}{2\tau_c}(e^{p_{\tilde{\xi}}}-e^{-p_{\tilde{\xi}}})-\frac{x}{\rho}(e^{-p_x}-1).
\end{eqnarray}

Similarly as in the negative binomial case, in the white-noise limit, $\tau_c\ll 1$, $\tilde{\xi}(t)$ and $p_{\tilde{\xi}}(t)$ equilibrate fast compared to $x$ and $p_x$. As a result, we can look for slowly-varying $x$ and $p_x$ dependent solutions of the third and fourth Hamilton equations~(\ref{HE}), which yields in the leading order of $\tau_c$
\begin{equation}\label{xieffOU}
\xi^{eff}=1-2(1-e^{-p_x})V x \tau_c.
\end{equation}
Note, that the value of $p_{\tilde{\xi}}^{eff}={\cal O}(\tau_c)\ll 1$ does not enter the equations for $\dot{x}$ and $\dot{p}_x$. Also, one can see that this result as well as the results below are independent of the arbitrary choice of $K$. Plugging this effective noise into the first of Hamilton equations~(\ref{HE}) we arrive at an effective 1D white-noise Hamiltonian
\begin{equation}
H(x,p_x)=f(x)(e^{p_x}-1)+x(e^{-p_x}-1)+x^2(e^{-p_x}-1)^2V\tau_c,
\end{equation}
which coincides with Eq.~(\ref{HredNoiseGamma}). As a result, in the white-noise limit the protein PDF under OU extrinsic noise coincides with the case of negative binomial EN. In particular, the variance in this case coincides with Eq.~(\ref{wnvargamma-main}). More generally, this indicates that in the white-noise limit, the choice of EN statistics does not affect the PDF in the leading order of $\tau_c\ll 1$.

In the adiabatic regime, $\tau_c\gg 1$, similarly as in the negative binomial case, we can use Eq.~(\ref{pdfA}) with $P(\xi)=1/\sqrt{2\pi\sigma_{ex}^2}\,e^{-(\xi-1)^2/(2\sigma_{ex}^2)}$, and
$P(n|\xi)=Ae^{-N\int_{g(\xi)}^x\ln\frac{y\xi}{f(y)}dy}$, where $x=g(\xi)$ solves the equation $f(x)=x\xi$. As a result, we arrive at Eq.~(\ref{Pnadia}), where here, the cost function satisfies
\begin{eqnarray}\label{fun}
\Phi(x,\xi)=\int_{g(\xi)}^x\ln\frac{y\xi}{f(y)}dy+\frac{(\xi-1)^2}{2V}.
\end{eqnarray}

Now, we use the fact that $N\gg 1$ and solve the integral in Eq.~(\ref{Pnadia}) via the saddle-point approximation. The saddle point is obtained at $\partial_{\xi}\Phi(x,\xi)=0$. Using Eq.~(\ref{fun}), this yields the following algebraic equation
\begin{equation}\label{der}
\frac{\partial \Phi}{\partial\xi}=\frac{x-g(\xi)}{\xi}+\frac{\xi-1}{V}=0.
\end{equation}
Solving the equation $V[x-g(\xi)]+\xi(\xi-1)=0$ for $\xi$ yields the optimal noise magnitude $\xi_*(x)$. Using this result and Eq.~(\ref{fun}), we find the PDF according to Eq.~(\ref{adiapdfgam-main}) in the main text. Note, that the resulting PDF here differs from the negative binomial case, since the cost function here [Eq.~(\ref{fun})] differs from that in the case of negative binomial EN [Eq.~(\ref{fungam})].

The variance of this PDF can be explicitly calculated by using Eq.~(\ref{secdergam}), with $x_*$ being the unperturbed fixed point $x_*=f(x_*)$. Since the variance is calculated in the close vicinity of the fixed point $x\simeq x_*$, similarly as for the negative binomial EN, we find the saddle point to be
\begin{equation}\label{xi}
\xi_*(x)\simeq 1+\frac{V[x-g(1)]}{Vg'(1)-1},
\end{equation}
which coincides with Eq.~(\ref{xigam}). Repeating the calculations in the same manner as in the case of negative binomial EN, we find
\begin{equation}\label{adiavar}
\sigma_{obs}^2=\frac{Nx_*}{1-f'(x_*)}\left[1+\frac{Vx_*}{1-f'(x_*)}\right].
\end{equation}
This result again coincides with the variance in the negative binomial case [Eq.~(\ref{adiavargam})]. This indicates that to determine the variance of the protein PDF under EN (in both the white- and adiabatic-noise limits), the complete statistics of the EN is less relevant. The only relevant parameter here is the width of the EN distribution, or its magnitude, given by the parameter $\sigma_{ex}$.

\setcounter{equation}{0}
\section{Correction of analytical variance using the numerical mean}
In cases where the EN magnitude is large, the mean of the distribution can shift, as discussed above. In these cases we apply a correction to the observed variance to account for the change in the IN. For example,  $\sigma_{obs}^2=N(1+V\tau_c)$ is corrected to
\begin{equation}
\sigma_{obs}^2 = \frac{\mu_{obs}}{N}\, N \, \left(1+\frac{\mu_{obs}}{N}\, V\tau_c\right),
\end{equation}
where $\mu_{obs}$ is the mean observed from numerical simulations, and $V=N\sigma_{ex}^2$.

\bibliography{references,references-ma}

\end{document}